# Enabling Adaptive Grid Scheduling and Resource Management

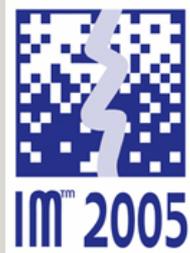

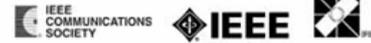

Aleksandar Lazarević
Dr Lionel Sacks, Dr Ognjen Prnjat

University College London

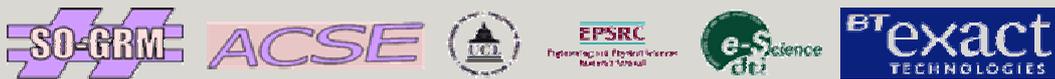

## Abstract:

Wider adoption of the Grid concept has led to an increasing amount of federated computational, storage and visualisation resources being available to scientists and researchers. Distributed and heterogeneous nature of these resources renders most of the legacy cluster monitoring and management approaches inappropriate, and poses new challenges in workflow scheduling on such systems. Effective resource utilisation monitoring and highly granular yet adaptive measurements are prerequisites for a more efficient Grid scheduler. We present a suite of measurement applications able to monitor per-process resource utilisation, and a customisable tool for emulating observed utilisation models. We also outline our future work on a predictive and probabilistic Grid scheduler. The research is undertaken as part of UK e-Science EPSRC sponsored project SO-GRM (Self-Organising Grid Resource Management) in cooperation with BT.

## Authors:

Aleksandar Lazarević    +44 20 7679 5755    a.lazarevic@ee.ucl.ac.uk
Dr Lionel Sacks         +44 20 7679 3976    lsacks@ee.ucl.ac.uk
Dr Ognjen Prnjat        +44 20 7679 5755    oprnjat@grnet.gr

# Introduction

- ## Problem Background
  - ### Non-automated deployment, maintenance and monitoring
  - ### Ineffective resource utilisation and scheduling
  - ### Lack of monitoring standard and interoperability with other components
- ## Methodology
  - ### Highly granular and adaptive resource monitoring $\Rightarrow$
  - ### Develop useful statistical models: job arrival & service rates, resource utilisation and execution time $\Rightarrow$
  - ### User-orientated, predictive and probabilistic scheduler

Creation of federated pools of computational and storage resources based on Grid technology has given scientists access to previously unreachable levels of processing power, within an open and community controlled environment. As more scientist from non-technology backgrounds seek to deploy Grid applications and use Grid platforms, the complexity of installation and maintenance of such systems has become all too obvious. Efforts have been undertaken to make Grid middleware more administration and user friendly, but it still requires highly specialised operators and managers.

Although some Grid pools contain thousands of CPUs over tens of different sites [1], effective and sustained utilisation is still dependant on manual tuning of multiple layers of schedulers and resource brokers. Most commonly used Grid schedulers have been adapted from legacy cluster systems, and while providing production-level reliability they are not flexible or adaptive enough to optimally harness Grid's distributed and heterogeneous resources.

With Virtual Organisations spanning many administration domains, monitoring both hardware fabric and Grid middleware and service becomes a daunting task. Current monitoring frameworks are capable of collecting large amount of measurements, yet they are rarely fed back into the management and scheduling structure where they could be of great value. The intrusion of monitoring system is often neglected, and little intelligence is built-in to adapt the frequency, scope and communication method to the current state of the Grid environment.

We recognise the need for a more granular and adaptive monitoring system in order to develop a better understanding of the nature and statistics of Grid workflows. Collected measurement data will initially be analysed off-line for statistical patterns of job arrival and service rates, resource utilisation and execution time. Based on these findings, a predictive and probabilistic scheduling algorithm will be developed for on-line analysis of historical data fed back from the monitoring and management tools.

# SO-GRM Overview

- ## Self-Organising Grid Resource Management
  - 3 year UK e-Science EPSRC sponsored project in collaboration with BT @ Adastral Park
  - Base research in autonomous Grid management

- ## Developed components:
  - SORD: Self-Organising Resource Discovery
  - I3: Autonomous Security Monitoring & Enforcement
  - SLAM: SLA Management and Negotiation

Self-Organising Grid Resource Management (SO-GRM) is a three year UK e-Science project founded by EPSRC, undertaken in industrial collaboration with BT research labs at Adastral Park.

The project's overall aim is a development of an autonomous management infrastructure able to support the job execution through its full lifecycle – from initial SLA admission control, through self-organising resource discovery over multiple domains, to run-time intrusion detection. The basic policy-based resource and security management concepts were developed within the context of a previous active networks project, and these components have been adapted and tested in a Grid scenario:

• Self-Organising Resource Discovery [2] is a resource discovery system based around a distributed protocol and autonomous agents running on each Grid node. It creates a small-world topology with each node linked to several topologically near neighbours and few random distant nodes. By using such network it is able to route resource queries to correct recipients based on limited local information. The protocol has been both simulated and deployed and has proved scalable and efficient.

• Integrity Information Intelligence (I3) [3] component is a distributed run-time intrusion detection system. It is a combination of anomaly and misuse detection able to distinguish suspicious resource utilisation patterns and classify them as reoccurring malicious activity or new and still ambiguous behavior. I3 uses mahalanobis classifiers in multidimensional space, and stores extracted feature sets locally. Anomaly descriptions are encoded in an XML antidote and the nodes are autonomously immunised by the system.

• SLA Management (SLAM) [4] system is a work-in-progress SLA admission control and negotiation component, currently still in development. It's primary objective is decomposition of higher-level Service Level Agreements into implicit resource requirements which can be used to negotiate and decide SLA acceptance and enforcement.

# SO-GRM Research Focus

- Further research focus:
  - A more granular and closely coupled monitoring system
  - Characterisation of Grid application load and development of parameterised load emulator
  - Predictive scheduler based on process' statistical properties

Relying on already developed components and group expertise in distributed system, our continuing research effort will focus on developing an autonomous statistical scheduling algorithm for Grid platforms.

Presented in this paper are two prerequisite components for such algorithm.

An improved monitoring system has been developed, able to measure individual CPU utilisation and memory footprint of each Grid submitted job. These metrics are fully integrated in standard monitoring data flow, and presented in a way suitable for off-line or on-line analysis. Operating from a common information base, all components including schedulers and resource brokers can access and use this additional information.

To facilitate testing and optimisation of new management components, a Grid application emulator tool has been developed. Resource utilisation patterns observed on production Grids running a variety of applications can thus be emulated on research testbeds.

We expect deployment of our monitoring system on UCL's newly installed Central Computing Cluster, a 200 CPU Grid environment on which more than 20 UK Grid projects will run. Collected measurements will be used to develop statistical understanding of Grid processes, and to accurately parameterise our load emulation tool for testing of our future scheduling algorithm.

# SO-GRM Architecture

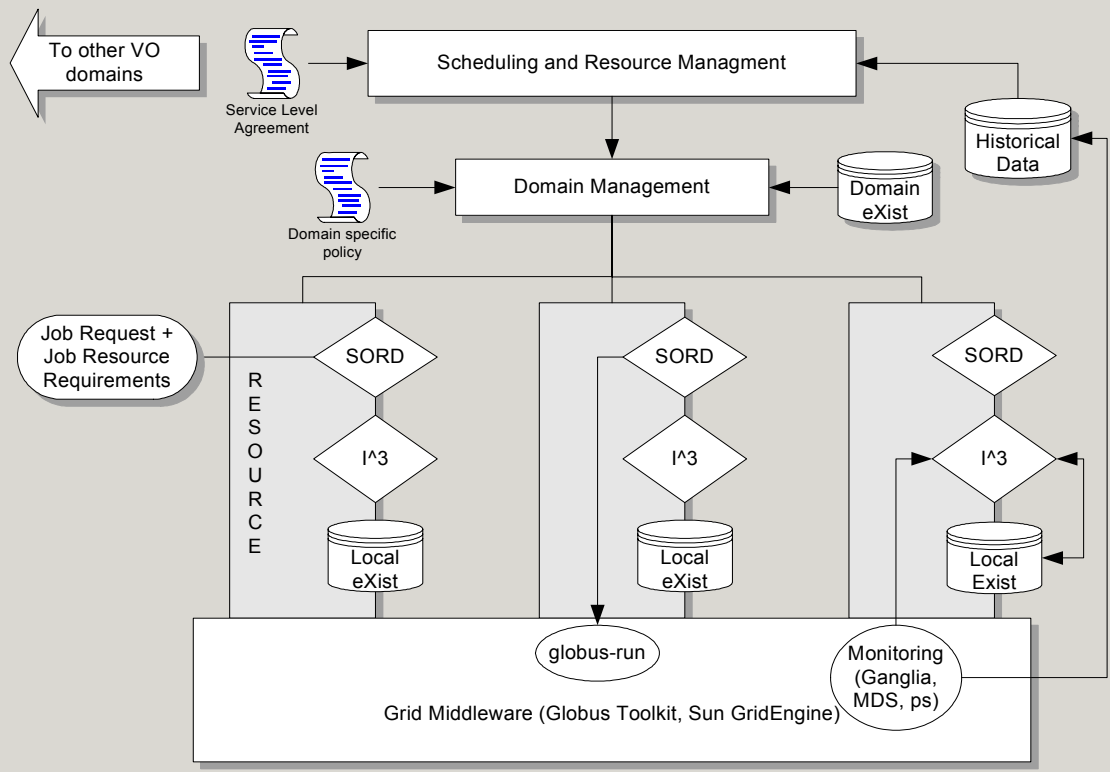

The figure presents the overall SO-GRM system architecture, as deployed on our multi-domain Grid testbed.

Our set-up was based on Globus Toolkit [5] version 2.4, but due to a modular design all components can support other Grid middleware, such as Sun Grid Engine [6] or Condor [7].

Each computational resource runs SORD and I3 agents together with a local XML database (eXist [8] or similar). Common monitoring system consolidates information provided by Ganglia monitoring system [9], Globus MDS [10], UNIX proc and our own custom information providers. These measurements are adapted in frequency and scope, and used by all management components, as further detailed in later sections.

Initially, an incoming service level agreement is parsed by SLAM, and based on historical loading data and current state of the local cluster a decision is made whether to support it or not. Future job submissions, tagged with the proper SLA id, can be directed to any SORD agent running in the cluster, removing the head or gateway node as a single point of failure. All Grid submitted processes are monitored by I3, and if anomalous behaviour is observed appropriate policies are executed.

SO-GRM architecture provides for SLA agreements between two or more Grid operators offering controlled and manageable overflow or failover capability. Once such SLA is admitted through SLAM, resource discovery component will be updated with topology shortcuts to other administrative domains and will be able to transparently discover available resources, provided SORD agents are running on remote nodes.

# Functional Demonstration

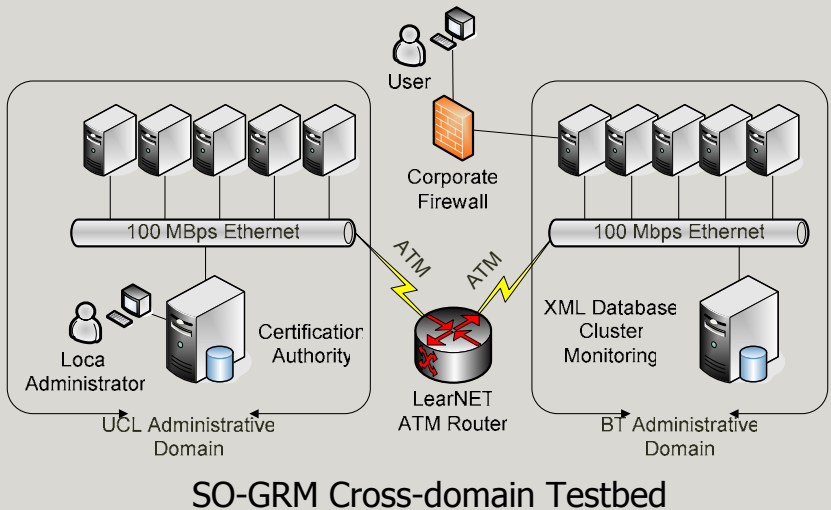

SO-GRM Cross-domain Testbed

Debug Console Screenshot

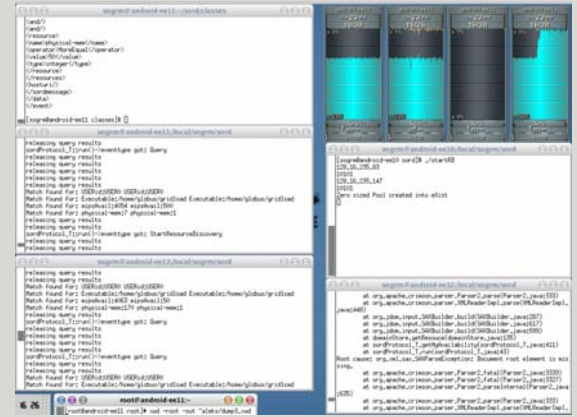

Component integration and functional testing was done on a testbed consisting of Linux RedHat 8 machines in UCL and BT administrative domains. Inter-domain link was through a routed ATM network. Each machine ran a full set of SO-GRM components and jobs were submitted to BT domain from a remote user. Policies were set to primarily assign jobs to BT resources, with overflow requests submitted to appropriate nodes in the UCL domain.

The screenshot shows several SORD debug windows and CPU and memory monitors for some of the machines in the UCL domain. Several types-of-service (ToS) were defined, therefore some service level agreements had exclusive usage of the resources, regardless of overall utilisation. As a result, certain machines were kept unloaded and awaiting high priority jobs, others were executing one user process at the time, while a subset was provided on a best-effort basis and had several concurrently executing process.

# Monitoring and Measurement

- Separation of volatile and non-volatile metrics
- Hierarchical distribution of monitoring data
- One data base adapted:
  - By measurement frequency
  - By Scope
  - By Consumer
- Data communication method adaptable:
  - Distributed $\Rightarrow$ effective
  - Centralised $\Rightarrow$ reliable

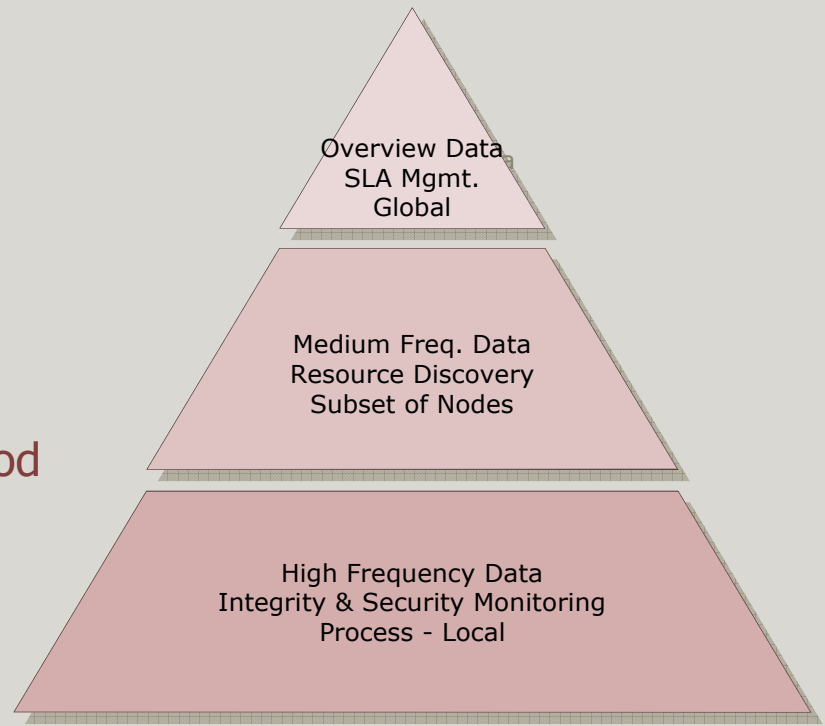

Overview Data
SLA Mgmt.
Global

Medium Freq. Data
Resource Discovery
Subset of Nodes

High Frequency Data
Integrity & Security Monitoring
Process - Local

SO-GRM monitoring framework operates in a hierarchical structure, with information communicated between the layers on a "need-to-know" basis. The core of the system is Ganglia cluster monitoring, chosen for its extensible interface, effective storage of data in fixed size round-robin databases, use of XML encoded measurements, and a customisable unicast/multicast delivery protocols. Our objective was to extend this system with custom information providers, monitoring sensors and management scripts that can provide detailed historical data to Grid scheduler as a basis of making prediction on future performance of computational resources.

Information on non-volatile, or rarely changing metrics, such as operating system version or hardware description is stored in Globus MDS using the GLUE [11] schema. Volatile metrics, such as CPU load, memory footprint or network bandwidth utilisation are sampled up to ten times a second on each Grid node, and this raw data is temporarily stored locally. l3 security subsystem uses these high-frequency, low latency measurements to examine the features of the processes and detect anomalies. A limited subset of this data is consolidated on one to five second basis and published throughout the local cluster. This information is used by the SORD resource discovery component to answer resource requests, or route them to the node most likely to be able to fulfil them. At the top of consumption pyramid, SLA management receives a low frequency overview snapshot of the Grid and each VO load, and tracks global temporal tendencies trying to predict the impact of new SLA requests on the overall system.

Data communication is also adapted to the nature of monitoring information. High frequency measurements are always kept local, avoiding the costs of transferring large amounts of raw data. Consolidated per-second data is broadcasted on the LAN, the system can tolerate loss of one or more packets (measuring points). SORD communicates effectively using a small-world topology and queries with limited TTL (time-to-live). At the overview level, the volume of the data is greatly reduced, and the requirements for complete and accurate capture require it to be kept in an appropriate relational database.

# Monitoring Functional Diagram

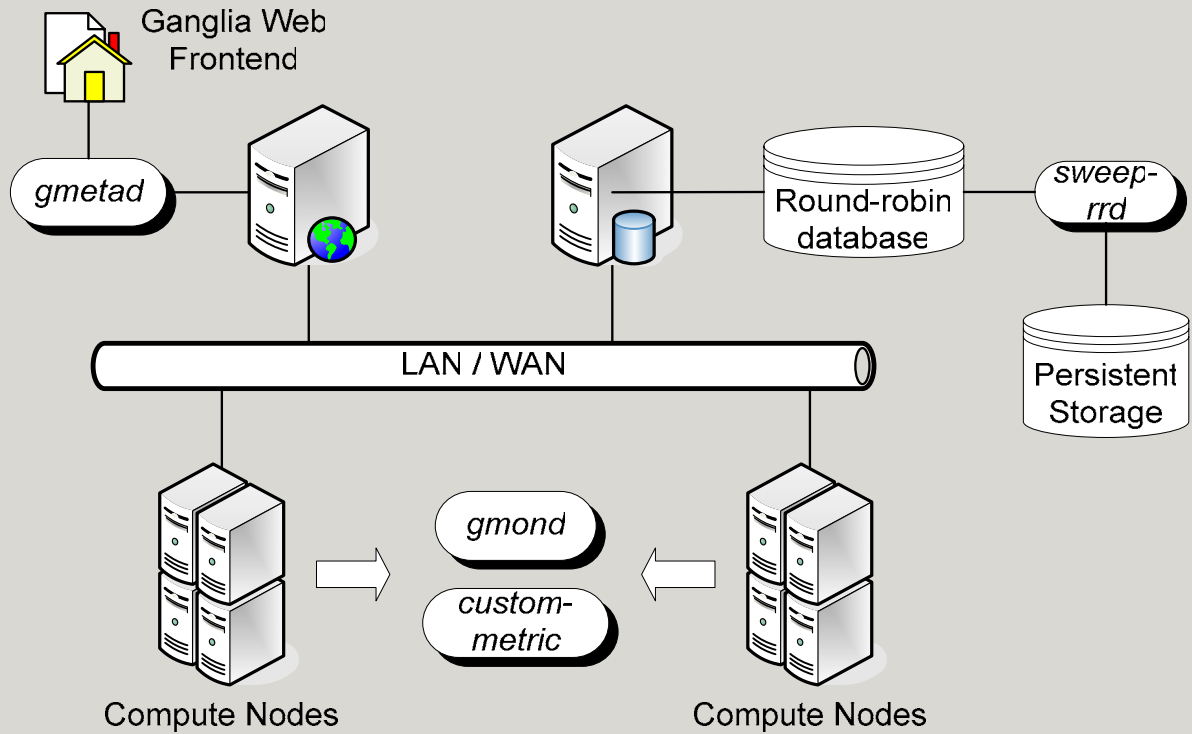

Integration of monitoring components is shown in this functional diagram. Each Grid node runs a Ganglia monitoring daemon that provides basic metric monitoring, and a custom SO-GRM monitor which offers a more granular monitoring of CPU utilisation and memory footprint of each Grid submitted process. Integration of these information providers is dependant on local administrative preference; a unified compiled application, or a combination of stock *gmond* and compiled or shell script SO-GRM monitor are all available. Each sampled measurement is broadcasted on the LAN within the cluster, and can be communicated in a point-to-point manner to any other IP address. One or more nodes in the Grid run Ganglia's *gmetad*, listening for broadcasts and storing data in round-robin databases (RRD). These operate as layered wheels storing decreasingly detailed time-series information. By using such hierarchical approach, database size remains constant but contains progressively more consolidated data from further in the past.

In order to extract and preserve highest frequency measurements, one or more nodes run SO-GRM database sweeper application which processes RRD files in regular time intervals before any consolidation has been done. Requested metrics and time ranges are recorded on persistent storage in the format appropriate for off-line analysis.

# Ganglia Monitoring – Screenshots

Ganglia Node Monitoring Screenshot

Ganglia VO Monitoring Screenshot

Ganglia provides a web interface to the measurements contained in the round-robin databases. Data is organised in a hierarchical manner starting with an overview of federation of clusters down to a detailed information on one specific node.

Screenshot in the top left presents the UCL VO during the UCL-BT cross-domain testing. Colour coding is used to represent different level of node utilisation, and overview graphs are displayed.

In the bottom right, numerous metrics pertaining to one node are displayed. SO-GRM monitoring of Globus submitted CPU utilisation is integrated in this view. In case of a space-shared resource, this per-grid-process CPU metric can offer insight in the performance the application is experiencing, and correlated with total machine load can point to poor application optimisation.

# Grid Application Load Simulation

- Scheduler testing problems:
  - Few simulation tools: SimGrid and GridSim
  - Little knowledge of Grid application statistics
  - Problems modelling dynamicity and heterogeneity of Grid resources
- Testing in self-organising and adaptive context:
  - Presented computational load must be variable and dynamic
  - Test runs must be controllable and repeated
  - Overall job arrival distribution must be controllable

While more Grids are being deployed as production environments, they often run few applications with similar resource utilisation profiles. As Grids establish themselves as multi-purpose computing platforms, more complex job statistics will develop. GridLoader application was developed out of need for a controllable and tuneable load generator that could be used to test other components of SO-GRM under realistic future utilisation scenarios. Testing of proof-of-concept software, such as SORD and I3, in a production environment is unlikely, yet deploying production software on research testbeds is either time consuming (from an administration point of view), or expensive (from software licensing perspective). Our planned development of statistical Grid meta-scheduler would eventually require similar testing, and would benefit greatly from optimisation in a realistic loading environment.

Our objective was not to create a scheduling simulator, like SimGrid [12] and GridSim [13], but to create a simple, parameterised resource loading application that can simulate job arrival and resource utilisation patterns observed using our monitoring system deployed in production Grid environment.

In order to fully test adaptability of our self-organising architecture, presented computational load was required to be dynamic and with variable statistical distribution. All experiments had to be repeatable, so that any changes in the management system which have led to improvement in performance could be registered. While these local probabilistic elements are required in order to present a dynamic load profile, the simulator needs to be controllable on global scale to enable us to test the impact of different overall job statistics.

# GridLoader – Functional Diagram

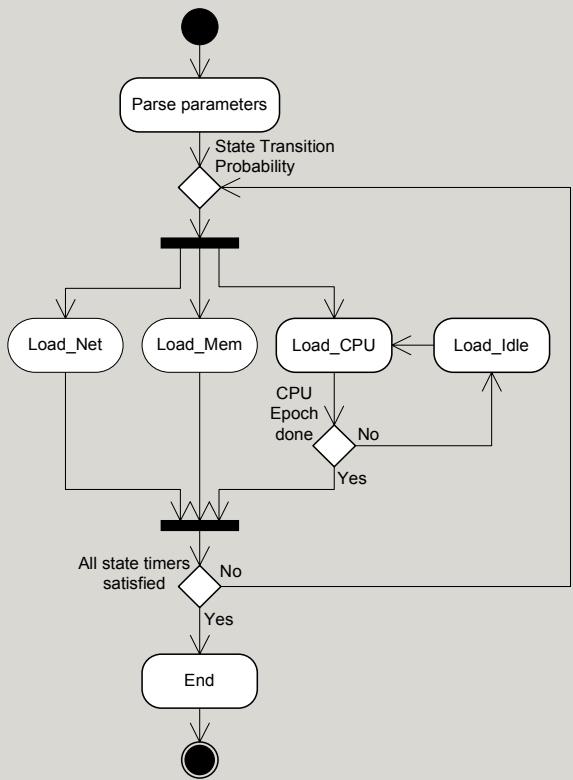

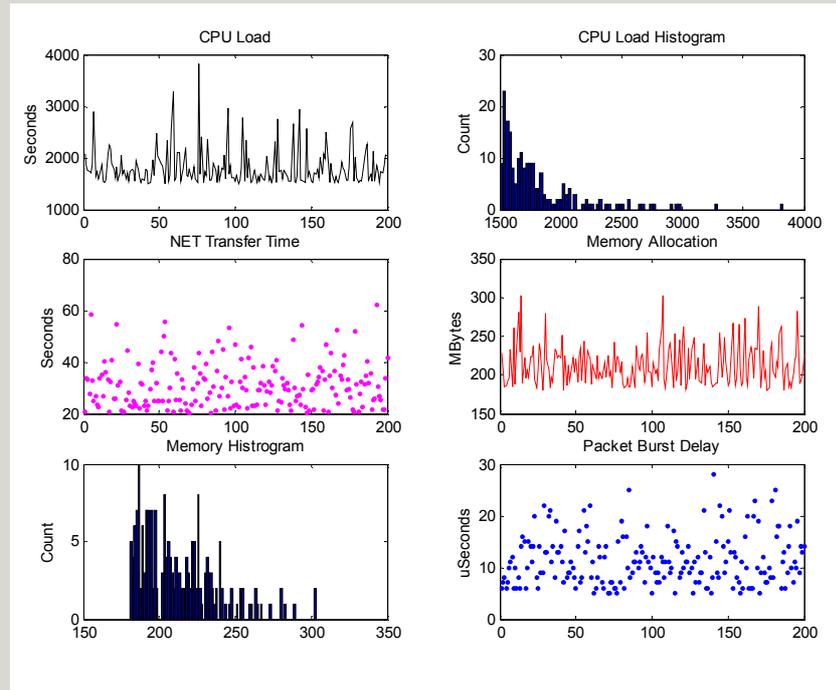

To ensure portability across different platforms GridLoader was developed in ANSI C without the need for any additional libraries. As shown on the UML Activity diagram, it is based on a state machine design, and able to stress CPU, memory and network subsystems. All aspects of component loading are parameterised either from a configuration file or at run-time using command line options. Network loading is characterised by the duration of network transfer and a bandwidth-setting inter-packet delay. Memory footprint is specified in megabytes, while CPU loading is expressed in the overall duration and a probabilistic factor influencing the level of CPU utilisation. These parameters give overall bounds and enable coordination of experiments across all the machines in the cluster (e.g. CPU loading time across all machines to follow a Pareto probability distribution function). However, each of the loading modules achieves its target in a probabilistic manner, so that the local monitoring facility will see different loading patterns even for equally parameterised runs.

An auxiliary component was developed in Matlab® to facilitate creation of parameter files for large job runs. Users can specify upper and lower bounds of each metric, and a probability function to be used for random number generation. A properly formatted GridLoader parameter file will be created, and values of all parameters will be plotted for visual inspection, as shown above right.

The current version of GridLoader uses a deterministic state transition table, progressing through network loading, memory allocation and CPU utilisation states in progression. This is similar to an "embarrassingly parallel" Grid application, such as parameter sweep tools, staging the input data, allocating required memory and proceeding to CPU intensive core calculations that would usually produce a small result data set. A more sophisticated model is possible when GridLoader is used with probabilistic state transition table where all three primary states are entered into many times with changing probabilities.

# GridLoader – Test Results

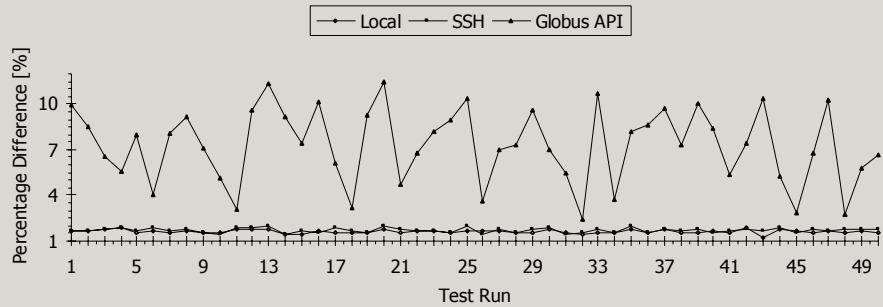

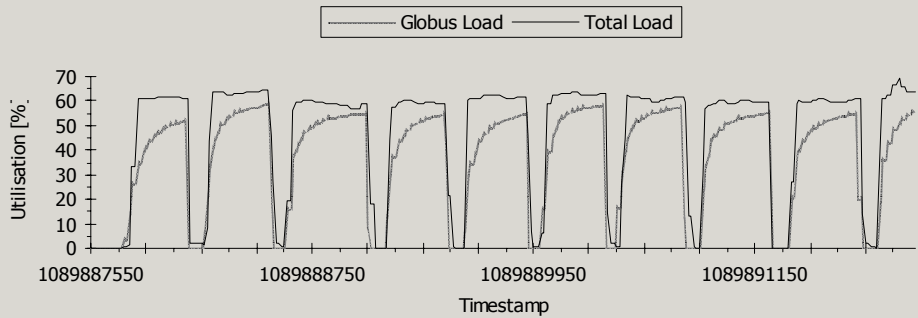

Reliability of GridLoader was tested by running a set containing 120 jobs taking around 24 hours to complete. The start and end time of each job was recorded and compared to expected execution times. The job set was first run locally, and then submitted to the same machine using two remote execution scenarios: Secure Shell (SSH) and Globus 2.4. The top figure shows the percentage difference in expected versus observed execution times for a subset of 50 jobs.

Running on the local node, actual GridLoader execution times are less then 2% greater than expected. This is due to system overhead needed in setting up network transfers and allocating memory, duration of which are not accounted in the timekeeping of the program. This operating-system intrinsic increase will be present for all applications, and we found that a realistic and accurate simulation of desired load can be achieved with GridLoader.

The bottom figure shows total CPU utilisation and Globus submitted GridLoader CPU utilisation over part of the test run. Data has been adjusted for the impact of monitoring components, although at complete idle they add less than 1% to the total machine load. The measurements capture the difference between GridLoader generated load and total system load which includes various background processes associated with Globus middleware, and kernel time servicing network transfers, memory allocation and process scheduling. On a space-shared node, it would be possible to independently track resource utilisation of each Grid job, thus providing more detailed information to the scheduler and management system.

# Adaptive Scheduler Research

- Lightweight, user transparent approach rules out
  - Application re-compilation or source code changes
  - Resource utilisation prototyping or instrumentation
  - Resource intensive prediction

- Develop a fast one-step-ahead prediction based on statistics of previous runs
- Exploit time and space locality of jobs and target machines
- Gradually improve confidence levels

The context of our research and developed architecture requires a lightweight, flexible and adaptive scheduler approach. As a result, we ruled out scheduling methods that would require application recompilation or re-linking, resource utilisation prototyping, loading instrumentation or predictions that would be computationally expensive. All these methods have previously been used with success in a limited number of applications or setups but fail to provide a universal and transparent solution. Instead, we will be developing a fast, one step ahead prediction engine based on the statistical data collected about previous runs. A feedback mechanism will gradually improve the quality of predictions and increase confidence levels.

As a starting point of our research, time and space locality of Grid workflow will be examined, together with the influence of social aspects, such as time of the day and submitting username. UCL's new Grid cluster with its widely varying user community will provide the bulk of monitoring data, although other Grid operators will also be approached.

# Conclusions

- Grid will require intelligent and autonomous managing components
- SO-GRM developed components:
  - SORD – small-world based distributed resource discovery
  - I3 – distributed intrusion detection
  - GridLoader – parameterised emulation of Grid applications load
  - Monitoring framework – highly granular and adaptive, integrated with other management components
- Further research in probabilistic Grid scheduler

University College London                                        A. Lazarević